\documentclass[acmsmall,screen,nonacm]{acmart}
\AtEndPreamble{\hypersetup{pdflang={en-US}}}
\usepackage{amsmath,booktabs,array}
\usepackage{tikz,pgfplots}
\pgfplotsset{compat=1.18}
\usetikzlibrary{arrows.meta,positioning}
\usepackage{placeins}
\usepackage{float}

\definecolor{figblue}{HTML}{3569A6}
\definecolor{figteal}{HTML}{218681}
\definecolor{figorange}{HTML}{BD6C2C}
\definecolor{figpurple}{HTML}{8060A5}
\definecolor{figgray}{HTML}{5D6672}
\usetikzlibrary{calc,fit,backgrounds,shapes.geometric}
\pgfplotsset{paperaxis/.style={axis line style={black!55},tick style={black!45},
 tick label style={font=\fontsize{8}{9}\selectfont},label style={font=\fontsize{8}{10}\selectfont},
 title style={font=\fontsize{9}{11}\selectfont},grid style={black!10},major grid style={black!10},
 legend style={font=\fontsize{8}{9}\selectfont,draw=none,fill=none},legend cell align=left,scaled ticks=false}}
\setcopyright{none}
\copyrightyear{}
\acmDOI{}
\makeatletter
\newcommand{\scope@author@affiliations}{%
  \raggedright
  \renewcommand\department[2][0]{##2,\space}%
  \def\city##1{\global\@ACM@citypresenttrue\unskip, ##1\ignorespaces}}
\pretocmd{\@mkauthors@i}{\scope@author@affiliations}{}%
  {\PackageError{SCOPE}{Cannot set author affiliation display}{Check the acmart class version.}}
\pretocmd{\@mkabstract}{\par\nobreak\hrule height0.4pt\relax\nobreak\smallskip}{}%
  {\PackageError{SCOPE}{Cannot set the abstract separator}{Check the acmart class version.}}
\makeatother

\author{Bolin Zhang}
\affiliation{%
  \department{National Supercomputing Center in Zhengzhou \& School of Computer and Artificial Intelligence}
  \institution{Zhengzhou University}
  \city{Zhengzhou}
  \country{China}}
\email{ieblzhang@gs.zzu.edu.cn}

\author{Zhiyuan Zhang}
\affiliation{%
  \department{National Supercomputing Center in Zhengzhou \& School of Computer and Artificial Intelligence}
  \institution{Zhengzhou University}
  \city{Zhengzhou}
  \country{China}}
\email{div1n352146@gs.zzu.edu.cn}

\author{Yongguang Lv}
\affiliation{%
  \department{National Supercomputing Center in Zhengzhou \& School of Computer and Artificial Intelligence}
  \institution{Zhengzhou University}
  \city{Zhengzhou}
  \country{China}}
\email{lvyongguang@gs.zzu.edu.cn}

\author{Ruiqing He}
\affiliation{%
  \department{School of Communication and Artificial Intelligence}
  \department{School of Integrated Circuits}
  \institution{Nanjing Institute of Technology}
  \city{Nanjing}
  \country{China}}
\email{njheruiqing@hotmail.com}

\author{Hengliang Guo}
\affiliation{%
  \department{National Supercomputing Center in Zhengzhou}
  \institution{Zhengzhou University}
  \city{Zhengzhou}
  \country{China}}
\email{guohengliang@zzu.edu.cn}

\author{Jiandong Shang}
\affiliation{%
  \department{National Supercomputing Center in Zhengzhou}
  \institution{Zhengzhou University}
  \city{Zhengzhou}
  \country{China}}
\email{sjd@zzu.edu.cn}

\author{Qiang Chen}
\affiliation{%
  \department{National Supercomputing Center in Zhengzhou}
  \institution{Zhengzhou University}
  \city{Zhengzhou}
  \country{China}}
\email{qiangchen@zzu.edu.cn}

\AtBeginDocument{\hypersetup{pdfauthor={Bolin Zhang, Zhiyuan Zhang, Yongguang Lv, Ruiqing He, Hengliang Guo, Jiandong Shang, Qiang Chen}}}
\authorsaddresses{\begingroup\urlstyle{same}
Authors' Contact Information: Bolin Zhang, National Supercomputing Center in Zhengzhou \& School of Computer and Artificial Intelligence, Zhengzhou University, Zhengzhou, China; e-mail: \href{mailto:ieblzhang@gs.zzu.edu.cn}{\nolinkurl{ieblzhang@gs.zzu.edu.cn}}; Zhiyuan Zhang, National Supercomputing Center in Zhengzhou \& School of Computer and Artificial Intelligence, Zhengzhou University, Zhengzhou, China; e-mail: \href{mailto:div1n352146@gs.zzu.edu.cn}{\nolinkurl{div1n352146@gs.zzu.edu.cn}}; Yongguang Lv, National Supercomputing Center in Zhengzhou \& School of Computer and Artificial Intelligence, Zhengzhou University, Zhengzhou, China; e-mail: \href{mailto:lvyongguang@gs.zzu.edu.cn}{\nolinkurl{lvyongguang@gs.zzu.edu.cn}}; Ruiqing He, School of Communication and Artificial Intelligence, School of Integrated Circuits, Nanjing Institute of Technology, Nanjing, China; e-mail: \href{mailto:njheruiqing@hotmail.com}{\nolinkurl{njheruiqing@hotmail.com}}; Hengliang Guo, National Supercomputing Center in Zhengzhou, Zhengzhou University, Zhengzhou, China; e-mail: \href{mailto:guohengliang@zzu.edu.cn}{\nolinkurl{guohengliang@zzu.edu.cn}}; Jiandong Shang (corresponding author), National Supercomputing Center in Zhengzhou, Zhengzhou University, Zhengzhou, China; e-mail: \href{mailto:sjd@zzu.edu.cn}{\nolinkurl{sjd@zzu.edu.cn}}; Qiang Chen (principal corresponding author), National Supercomputing Center in Zhengzhou, Zhengzhou University, Zhengzhou, China; e-mail: \href{mailto:qiangchen@zzu.edu.cn}{\nolinkurl{qiangchen@zzu.edu.cn}}.
\endgroup}

\title[SCOPE: Physical Execution Latency-Guided Compilation]{SCOPE: A Physical Execution Latency-Guided Quantum Circuit Compilation Framework}
\makeatletter
\let\scopebibliography\thebibliography
\renewcommand{\thebibliography}[1]{\scopebibliography{#1}\clubpenalty=0\@clubpenalty=0\widowpenalty=0\interlinepenalty=0}
\makeatother
\begin{document}
\begin{abstract}
Quantum computing requires compilers to translate algorithms represented in the quantum circuit model into unitary operations supported by physical devices. Limited coherence time makes physical execution latency an important compilation objective. Gate fusion can shorten pulses, but gate count and depth do not capture fused targets' control demands. Evaluating every candidate with quantum optimal control (QOC) would embed pulse optimization in structural search. We present Schmidt-guided Compilation for Optimized Pulse Execution (SCOPE), driven by an operator Schmidt time surrogate. The target's angular distance from product operators measures the required nonlocal change; dividing it by a coupling-limited evolution-rate bound gives a lower bound on exact implementation time. Trace-angle and waveform constraints complete block scores for critical-path evaluation during genetic search. Local QOC determines final pulse durations. Under a numerical nearest-neighbor control model, all 104 results across 13 instances from four application families have shorter QOC critical-path durations than an adapted efficient pulse generation framework with advanced synthesis for quantum circuits (EPOC). Using eight-run instance means, the equally weighted reduction is 42.50\%, with a geometric mean speedup of 2.00. In a shared six-qubit quantum Fourier transform (QFT) ablation, time-surrogate guidance reduces mean QOC duration by 22.21\% relative to gate-count guidance and 21.01\% relative to depth guidance, validating the surrogate's joint use in fitness evaluation and mutation-region selection.
\end{abstract}


\maketitle
\section{Introduction}
\label{sec:introduction}

Quantum computing uses controllable quantum systems to address tasks such as quantum simulation and combinatorial optimization. Quantum simulation probes many-body states and dynamics in materials science, chemistry, and high-energy physics~\cite{Daley2022Simulation}. Quantum optimization examines where quantum algorithms offer benefits in solution quality and computational resources as problem size grows~\cite{Abbas2024Optimization}. Both depend on sufficient qubit resources and accurate implementation of the required evolution.

Hardware and error-control advances are improving this capability. Superconducting surface-code memories have demonstrated operation below the error-correction threshold, with logical error rates decreasing as code distance increases~\cite{Google2025BelowThreshold}. Such progress relies on precise physical control. On pulse-driven devices, logical operations become continuous-time evolutions constrained by drive amplitudes, bandwidth, and couplings; the control fields affect both duration and error~\cite{Koch2022QOC}. A compiler must therefore connect logical circuit choices to physical execution time.

Verified rewriting and learned transformation or pass selection explore alternative gate sequences~\cite{Xu2022Quartz,Li2024Quarl,Xu2025GUOQ,Quetschlich2025MQTPredictor}, while tensor, ZX, and Pauli representations support algebraic simplification~\cite{Ruiz2025AlphaTensor,Duncan2020ZX,Li2022Paulihedral,Paykin2023PCOAST}. Block synthesis generates alternative decompositions~\cite{Kukliansky2023QFactor,Smith2023LEAP,Patel2022QUEST}, and permutation-aware synthesis and mapping account for gate costs and inter-block communication~\cite{Liu2023PAS}. Evaluating these choices for physical latency requires knowing the control demands of the resulting targets.

Gate fusion makes these demands explicit: it combines a local sequence into one target unitary for quantum optimal control (QOC), which optimizes pulses across the original gate boundaries~\cite{Koch2022QOC,Khaneja2005GRAPE}. The pulse realizes the fused target unitary, so the durations assigned to its constituent gates do not determine the fused block's duration. Different sequences can produce the same target, and equal-length sequences can produce targets with different rotation and interaction requirements. A timing criterion must distinguish these requirements. At circuit level, dependencies and qubit occupancy determine which operations must wait for one another~\cite{Basilewitsch2024CircuitQSL}. Latency evaluation must therefore connect the control demands of the fused targets to their position in the executable schedule.

Pulse-aware methods obtain timing information through calibrated-pulse scaling and reusable control mappings~\cite{Earnest2021Pulse,Preti2022PulseMeta}, causality-aware grouping of QOC targets~\cite{Lee2026DArQ}, and direct compilation from target matrices to controls~\cite{Cho2024DirectPulse}. These methods inform structural search when their timing information is available while candidate targets can still change.

PAQOC and EPOC expose the remaining gap in evaluating new targets. PAQOC ranks merges by their effect on the critical path; in part of this analysis, merges that enlarge the qubit support are evaluated using average gate latency at that width~\cite{Chen2023PAQOC}. This estimate does not distinguish the control requirements of different targets of equal width. EPOC guides block synthesis by unitary distance, then regroups the result for QOC~\cite{Cheng2025EPOC}. The distance measures synthesis accuracy; different decompositions can achieve the same accuracy while producing different pulse durations. Structural search therefore needs a target-dependent timing criterion that distinguishes these control requirements before solving for pulses.

A physical time lower bound supplies this criterion by quantifying the evolution a target necessarily requires under bounded controls. Nonlocal resource measures and geometric speed-limit analyses connect target geometry to this evolution~\cite{Nielsen2003Dynamics,Impens2025GeometricControl}. Sampling and bandwidth constraints further shape achievable durations~\cite{Dionis2023SampledControl,Muller2022InformationTime}. A lower bound can be evaluated from these ingredients before any waveform has been chosen. This makes it available at the point where the compiler still controls the target operators and their dependencies. Expressing each block's necessary evolution in time units also lets the compiler combine local control demands along a dependency path. The resulting objective can compare layout, gate order, and routing through their effect on a common physical quantity: the circuit's required execution time.

We introduce SCOPE (Schmidt-guided Compilation for Optimized Pulse Execution). The operator Schmidt spectrum is invariant under unitaries acting separately on either side of a cut~\cite{Nielsen2003Dynamics}; its largest normalized coefficient determines the target's angular distance from product operators. This distance measures the nonlocal change requiring cross-cut coupling. Dividing it by the coupling-limited evolution rate gives a necessary time for exact implementation. SCOPE combines this bound with trace-angle and waveform constraints, then schedules the block scores to obtain a circuit-level time surrogate.

The surrogate guides a genetic algorithm (GA) over operation priorities, initial layouts, and routing weights. Each candidate is decoded, fused, and scheduled before scoring; selected solutions undergo QOC and rescheduling to determine physical execution time.

SCOPE links physical timing information to shorter final execution through three contributions:
\begin{itemize}
 \item A target-dependent time score available before pulse optimization. Operator Schmidt geometry and coupling-limited evolution give gate-time lower bounds, complemented by trace-angle and waveform constraints.
 \item Joint search directed at circuit latency. Mixed-encoding genetic search uses surrogate critical-path evaluation for fitness and mutation-region selection, linking layout, gate order, and routing to post-fusion timing.
 \item Lower final QOC latency and evidence for the surrogate's effectiveness. Across 13 instances, all 104 SCOPE results have shorter QOC critical-path durations than adapted EPOC~\cite{Cheng2025EPOC}, with an equally weighted mean reduction of 42.50\%. In the shared QFT-6 pipeline, time-surrogate guidance reduces mean QOC duration by 22.21\% relative to gate-count guidance and 21.01\% relative to depth guidance.
\end{itemize}

\section{Motivation}
\label{sec:motivation}

\subsection{Timing Fused Targets before Pulse Optimization}

Physical execution latency depends on the time needed to realize the operations that a compiler produces. Native-gate durations support gate-by-gate scheduling, but fusion replaces a sequence of gates with a single target unitary. Its control time is determined by the full product operator under the Hamiltonian, amplitude, and waveform constraints. QOC finds a pulse satisfying these conditions~\cite{Koch2022QOC}; candidate evaluation needs access to target-dependent timing information before repeatedly solving that control problem.

This requirement also arises at circuit level. Dependencies and physical-qubit occupancy define a path set $\mathcal P$, and block durations $t_b$ give completion time $T=\max_{p\in\mathcal P}\sum_{b\in p}t_b$. Layout, routing, and gate order change both the fused targets and these paths. A useful structural-search score must therefore connect the timing requirements of each new target to the completion time of the candidate circuit.

PAQOC's width-based timing estimates omit variation among equal-width targets~\cite{Chen2023PAQOC}, whereas EPOC obtains target-specific durations through QOC after synthesis and regrouping~\cite{Cheng2025EPOC}. Comparing new structural candidates therefore requires either resolving this variation before pulse optimization or repeatedly solving the control problem. SCOPE takes the former approach by computing necessary timing constraints from the target and physical resources.

Gate-time lower bounds provide such a criterion. Under fixed control constraints, let $t^\star(U)$ be the minimum time for exact implementation of target $U$, and let $t_{\mathrm{feas}}(U)$ be the duration of any exact implementation. A lower bound $L(U)$ satisfies
\begin{equation}
 L(U)\le t^\star(U)\le t_{\mathrm{feas}}(U).
 \label{eq:time_bracket}
\end{equation}
The bound describes a necessary time cost imposed by the target and available controls. This information is useful before pulse optimization because structural search chooses among target operators while their controls are still unknown. Each candidate changes the target operators produced by fusion and the paths on which they execute. Evaluating them under the same control limits makes their necessary timing requirements comparable. Search can then favor a target that demands less cross-cut evolution or a schedule that places fewer such demands in series. QOC subsequently constructs pulses for the selected targets. A lower bound connects these stages through physical resources fixed before the structural decision. We use the operator Schmidt spectrum to obtain this information directly from the target.

\begin{figure}[htbp]
\centering
\includegraphics[width=\linewidth]{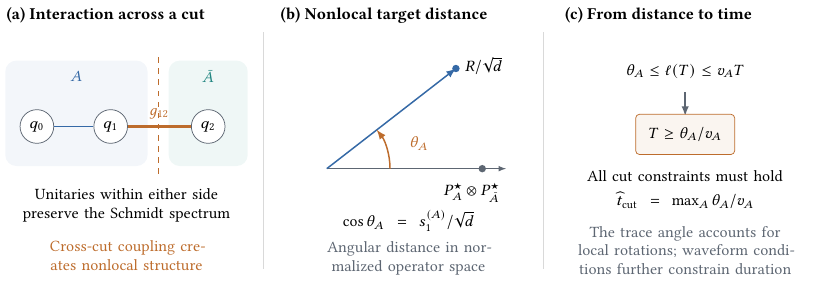}
\caption{How target structure constrains control time. (a) Unitary operations within either side preserve the Schmidt spectrum; cross-cut coupling creates nonlocal structure. (b) $P_A^\star\otimes P_{\bar A}^\star$ is a normalized product operator closest to the normalized target, and $\theta_A$ is their angular distance. (c) Control amplitudes bound the evolution path length $\ell(T)$ and hence the required time. Taking the maximum enforces the constraints of all cuts. The geometry is schematic.}
\label{fig:mechanism}
\Description{Three-panel mechanism diagram. A highlighted coupling crosses a bipartition. An angle represents the distance from a target operator to product operators. Arrows connect this distance and a maximum evolution rate to a lower bound on duration.}
\end{figure}

The reason for choosing this spectrum is that it distinguishes evolution requiring interaction across a qubit partition. Unitary operations confined to either side preserve the operator Schmidt spectrum~\cite{Nielsen2003Dynamics}. Since the identity is a product operator, generating a target with nonzero distance from the product-operator set requires cross-cut coupling. The largest normalized Schmidt coefficient determines this angular distance, $\theta_A$, and hence quantifies the nonlocal change required by the target (Fig.~\ref{fig:mechanism}).

Control amplitudes then convert this geometric requirement into a time bound. In an interaction picture that removes evolution within each side, coupling limits give a maximum speed $v_A$. A path reaching the target must cover at least $\theta_A$, while its length cannot exceed $Tv_A$; therefore $T\ge\theta_A/v_A$. The same pulse must satisfy every cut constraint, so their maximum remains a necessary duration. Local rotations can require time even when all cut angles vanish, which motivates the trace-angle term. Waveform constraints further account for the admissible pulse envelope and time grid. Section~\ref{sec:surrogate} combines these requirements into the block score.

This separation of target geometry and control resources also explains the bound's dependence on coupling strength. Scaling every allowed coupling across a cut by a positive factor $\alpha$ leaves $\theta_A$ unchanged and scales $v_A$ by $\alpha$. The cut contribution $\theta_A/v_A$ therefore scales by $1/\alpha$. The angle specifies the nonlocal change required by the target, and the speed bound specifies how quickly the available couplings can generate it. Under a common control model, candidate targets are compared using this same conversion from geometric change to physical time.

\begin{figure}[htbp]
\centering
\includegraphics[width=\linewidth]{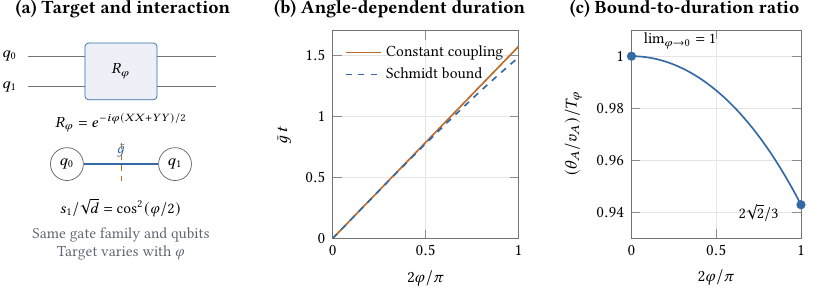}
\caption{Time bound and feasible duration for the XY gate family under a coupling-amplitude constraint alone. (a) Varying the interaction angle $\varphi$ leaves the gate type and participating qubits unchanged but changes the target Schmidt coefficient. (b) The cut bound and the feasible duration under constant coupling both vary with the target angle. (c) The bound-to-duration ratio approaches 1 in the small-angle limit and equals $2\sqrt2/3$ at $\varphi=\pi/2$. Appendix~\ref{app:xy} gives the derivation.}
\label{fig:xy_bound}
\Description{Three-panel analytic example. A two-qubit XY target is followed by curves of the lower bound and constant-coupling duration versus interaction angle. Both increase with angle. Their ratio decreases from the small-angle limit of one to approximately 0.943.}
\end{figure}

The XY family in Figure~\ref{fig:xy_bound} shows how this bound responds to target-dependent timing. Varying the interaction angle leaves gate type and participating qubits unchanged, but changes both the Schmidt angle and the duration under constant coupling. The ratio of the time bound to this feasible duration approaches one at small angles. The bound thus recovers a control-time scale within one gate family from target geometry and coupling strength.

Each bipartition of a fused target yields a necessary duration bound; their maximum therefore combines the cut constraints. The block score takes the larger of this cut bound and the trace-angle bound, then incorporates the waveform constraints and time grid. At circuit level, blocks along a dependency path execute sequentially. Summing their block scores along each path and taking the largest path sum gives a circuit-level bound for exact execution under the fixed decomposition and dependency graph.

With a fixed block-width limit, evaluating the bound requires matrix realignment, the largest singular value, and a trace, making repeated timing evaluation possible without waveform propagation and gradient iterations. For a fixed partition, critical-path duration is monotone in block durations, so scheduling these bounds also gives a circuit-level bound. We use this target- and resource-dependent quantity to guide structural search, then evaluate the selected implementations with QOC.

\subsection{Joint Choices of Layout, Gate Order, and Routing}

The time surrogate provides a way to compare candidates; search must then explore the structural choices that generate them. Changing a ready gate's priority changes the next operation decoded, which can affect routing swaps and subsequent fusion. The outcome also depends on the initial layout, so layout, gate order, and routing need to be searched jointly. For $n$ logical qubits, the initial layout alone has $n!$ permutations, and gate-order and routing choices further enlarge the search space.

These choices interact through the targets and dependencies created by the decoder. A layout determines which interactions need routing, while gate order determines when those routing operations enter the sequence and which operations can subsequently fuse. Their combined timing effect therefore becomes visible in the decoded block graph. Evaluating complete candidates retains this interaction in a single objective: a change is favorable when it reduces the longest weighted path after decoding and fusion. The search representation can then separate the decision variables without separating their physical evaluation.

A candidate can be scored only after gate ordering, routing, and fusion. Discrete choices in this process create jumps or intervals of unchanged fitness as priorities, layouts, and routing weights vary. A GA can use these complete-candidate scores directly to compare structural combinations. We therefore encode the layout permutation together with numerical vectors for operation priorities and routing weights, using crossover and mutation suited to each representation. Each new individual passes through the same decoding and surrogate evaluation, and its score determines parent selection and retention. Section~\ref{sec:search} gives the encoding, decoding, and population-update procedures.

\section{Framework Design}
\label{sec:design}

\subsection{Overview and Problem Formulation}
\label{sec:overview}

\begin{figure}[htbp]
\centering
\includegraphics[width=\linewidth]{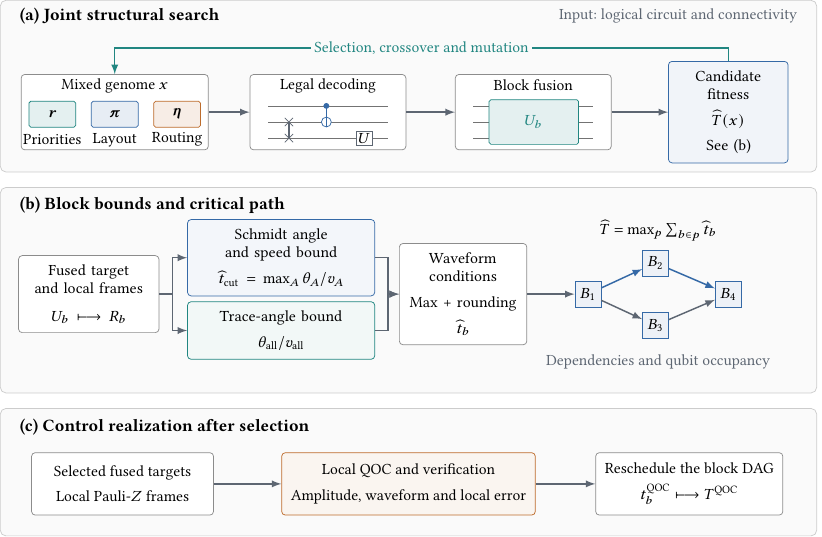}
\caption{SCOPE workflow. (a) Genetic search updates priorities, layouts, and routing weights before decoding and fusion. (b) Operator geometry and control bounds give block scores for critical-path evaluation. (c) Local QOC and independent numerical verification supply pulse durations for final rescheduling. Circuits and dependency graphs illustrate these stages.}
\label{fig:framework}
\Description{Workflow linking candidate generation, decoding and fusion, surrogate timing, and local pulse optimization of selected solutions. Search-stage scores feed back to the genetic algorithm. Feasible block durations enter the dependency graph for final rescheduling.}
\end{figure}

SCOPE decodes candidates comprising operation priorities, an initial layout, and routing weights into physical operations, then fuses them into contiguous local blocks (Figure~\ref{fig:framework}). Surrogate block durations give the candidate's critical-path score. QOC supplies verified pulse durations for rescheduling the selected solution on the same dependency graph.

Let the fused dependency graph be $\mathcal D=(\mathcal B,\mathcal E)$, where block $b\in\mathcal B$ has duration $t_b$. Completion times are computed in topological order as
\begin{equation}
 f_b=t_b+\max_{a\in\operatorname{pred}(b)} f_a,
 \qquad T(\mathcal D,\boldsymbol t)=\max_{b\in\mathcal B}f_b,
 \label{eq:schedule}
\end{equation}
Here $\operatorname{pred}(b)$ contains the immediate predecessors of $b$, and an empty maximum is zero. Dependencies and qubit occupancy constrain parallel execution. Decoding and fusion produce $\mathcal D(x)$ subject to connectivity, block-width, and contiguity constraints. Search minimizes $\widehat T(x)=T(\mathcal D(x),\widehat{\boldsymbol t})$; final evaluation gives $T^{\mathrm{QOC}}(x)=T(\mathcal D(x),\boldsymbol t^{\mathrm{QOC}})$ using verified pulse durations.

\subsection{Operator Schmidt Time Surrogate}
\label{sec:surrogate}

\subsubsection{Control Model and Local Frames}

The surrogate bounds exact-target implementation time using cut distances for nonlocal evolution and a trace angle that also captures local rotations. On a nearest-neighbor chain, each fused block contains at most three contiguous qubits, with local drives and XY couplings:
\begin{equation}
 \frac{H(t)}{\hbar}=\frac12\sum_i\bigl[u_{ix}(t)X_i+u_{iy}(t)Y_i\bigr]
 +\frac12\sum_{e=(i,i+1)} g_e(t)\bigl[X_iX_{i+1}+Y_iY_{i+1}\bigr].
 \label{eq:hamiltonian}
\end{equation}
The model has zero drift and sets couplings between distinct parallel blocks to zero. Local drives and couplings obey $|u_{i\alpha}(t)|\le\Omega_i$ ($\alpha=x,y$) and $|g_e(t)|\le\bar g_e$, respectively, where $\Omega_i$ and $\bar g_e$ are the corresponding angular-frequency amplitude bounds. The waveforms undergo finite impulse response filtering and piecewise-linear interpolation while satisfying endpoint conditions. Blocks that share no physical qubits and satisfy the dependency constraints can execute independently in parallel and are scheduled according to Eq.~\eqref{eq:schedule}.

For the target unitary matrix $U$ of a $k$-qubit block, let $d=2^k$. A discrete local frame is a tensor product of single-qubit $I$ or Pauli $Z$ operators. We choose
\begin{equation}
 G^\star\in\underset{G\in\{I,Z\}^{\otimes k}}{\arg\max}\,
 |\operatorname{Tr}(GU)|,\qquad R=G^\star U.
 \label{eq:frame}
\end{equation}
The incoming frame $F\in\{I,Z\}^{\otimes k}$ and residual operator $R$ jointly determine the physical control target. Discrete frame updates carry zero duration in this model. Define
\begin{equation}
 P=FRF,\qquad F_{\mathrm{out}}=G^\star F,\qquad U=F_{\mathrm{out}}PF,
 \label{eq:frame_transport}
\end{equation}
Physical control targets $P$, with input and output frames passed between blocks. The choice in Eq.~\eqref{eq:frame} preserves the operator Schmidt spectrum across each bipartition and minimizes the trace angle over the allowed frame set. Conjugation by $F$ changes only the signs of local drives and XY couplings, preserving symmetric amplitude bounds and waveform constraints under linear filtering. Since $\operatorname{Tr}(FRF)=\operatorname{Tr}R$ and the operator Schmidt spectra are also unchanged, the surrogate duration evaluated on $R$ is independent of the incoming frame.

\subsubsection{Operator Distances and Evolution Rates}

For each distinct nontrivial bipartition $A|\bar A$, realign the input and output indices of $R$ as
\begin{equation}
 [M_A]_{(a,a'),(b,b')}=R_{(a,b),(a',b')},\qquad
 M_A\in\mathbb C^{4^{|A|}\times4^{k-|A|}},
 \label{eq:realignment}
\end{equation}
where $a,a'$ and $b,b'$ label computational-basis states of the two subsystems. With Hilbert--Schmidt-orthonormal local operator bases, the operator Schmidt coefficients $s_j^{(A)}$ satisfy $\sum_j(s_j^{(A)})^2=d$; the largest normalized spectral weight and its angle are
\begin{equation}
 \lambda_A=\lambda_{\max}\!\left(\frac{M_AM_A^\dagger}{d}\right)
 =\frac{(s_1^{(A)})^2}{d},\qquad
 \theta_A=\arccos\sqrt{\lambda_A}.
 \label{eq:schmidt}
\end{equation}
The spectral weight $\lambda_A$ varies continuously with the target. For physical coupling edges $\delta A$ crossing the cut, define
\begin{equation}
 v_A=\sqrt{\frac12\sum_{e\in\delta A}\bar g_e^2},
 \qquad \widehat t_{\mathrm{cut}}=\max_A\frac{\theta_A}{v_A}.
 \label{eq:cut}
\end{equation}
Here $\theta_A$ is the angular distance from the normalized target operator to the set of product operators. Unitary operations acting separately on either side preserve this distance, while coupling across the cut allows evolution away from the product-operator set. Removing the evolution within each side through an interaction picture gives the speed bound $v_A$ from the control-amplitude limits. Every exact implementation therefore satisfies $T\ge\theta_A/v_A$. All cut constraints must hold simultaneously, so we take their maximum. For a single-qubit block, $\widehat t_{\mathrm{cut}}=0$. Appendices~\ref{app:geometry} and~\ref{app:resources} give the derivation.

The cut angles do not distinguish local rotations within product unitaries. To account for their time requirements, the surrogate also uses the trace angle between the residual operator $R$ and the identity. Combining the local drive and coupling bounds, we define
\begin{align}
 \theta_{\mathrm{all}}&=\arccos\frac{|\operatorname{Tr}R|}{d},
 &v_{\mathrm{all}}&=\sqrt{\frac{\sum_i\Omega_i^2+\sum_e\bar g_e^2}{2}},\\
 \widehat t_{\mathrm{raw}}&=\max\left\{\widehat t_{\mathrm{cut}},
                  \frac{\theta_{\mathrm{all}}}{v_{\mathrm{all}}}\right\}.
 \label{eq:raw}
\end{align}

\subsubsection{Waveform Constraints and Timing Score}

Exact implementation must satisfy both geometric constraints, giving $T\ge\widehat t_{\mathrm{raw}}$. Waveform constraints strengthen this bound.

Let $\Delta t$ be the time-grid spacing and $\tau_{\min}$ the shortest nonzero waveform duration. The control channels share a nonnegative, normalized FIR kernel, zero endpoint padding, and piecewise-linear interpolation. These conditions define a common normalized amplitude envelope with area $T-\tau_{\mathrm{pad}}$, where $\tau_{\mathrm{pad}}$ is the fixed area deficit relative to an ideal rectangular envelope, expressed in units of time. Combining the envelope constraint with the time grid gives the block surrogate
\begin{equation}
 \widehat t_b=
 \begin{cases}
 0,&R\text{ is the identity operator up to a global phase},\\
 \Delta t\left\lceil\dfrac{\max\{\tau_{\min},\tau_{\mathrm{pad}}+
                       \widehat t_{\mathrm{raw}}\}}{\Delta t}\right\rceil,
 &\text{otherwise}.
 \end{cases}
 \label{eq:proxy}
\end{equation}

Appendix~\ref{app:waveform} derives the envelope-area constraint and rounding to the time grid. For a fixed block partition and dependency graph, critical-path duration is nondecreasing in each block duration, so Eq.~\eqref{eq:schedule} uses the block bounds to score the circuit. Search uses scores constructed from exact-target bounds, and the selected solutions undergo control optimization at the block accuracy specified in Section~\ref{sec:experiment_settings}.

\subsection{Surrogate-Guided Structural Search and Fusion}
\label{sec:search}

Each individual specifies a complete set of compilation choices. For a circuit with $m$ logical operations, we write
\begin{equation}
 x=(\boldsymbol r,\pi,\boldsymbol\eta),
 \label{eq:genome}
\end{equation}
where $\boldsymbol r$ is the operation-priority vector, $\pi$ is the initial logical-to-physical mapping, and $\boldsymbol\eta$ contains routing-cost weights. Priorities affect gate order through the decoder, which selects the next operation only among those whose dependencies are satisfied and that can execute under the current layout. If ready two-qubit gates are not adjacent, the router selects a nearest-neighbor swap using costs that include the current operations and look-ahead distances, then decoding continues. Both routing swaps and the operations needed to restore the final mapping are included in the physical circuit.

\begin{figure}[htbp]
\centering
\includegraphics[width=\linewidth]{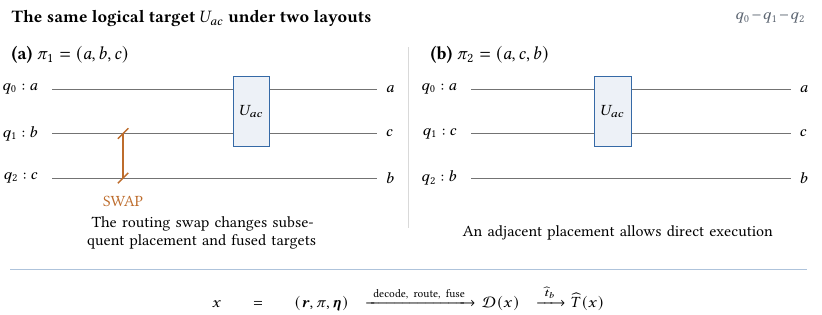}
\caption{Layout-dependent routing for target $U_{ac}$. The first layout requires a nearest-neighbor swap; the second permits direct execution. Full-circuit evaluation includes all operations and final mapping restoration.}
\label{fig:search}
\Description{Two three-qubit circuits compare layouts. The ordering a, b, c requires an adjacent swap of b and c before executing a target on a and c. The ordering a, c, b executes that target directly. Arrows below connect layout selection to decoding, fusion, and critical-path scoring.}
\end{figure}

Layout can eliminate a routing swap for the same target (Figure~\ref{fig:search}), but also changes subsequent operation locations. Joint encoding lets crossover and mutation explore these coupled choices, whose effect is evaluated after fusion and scheduling.

Greedy fusion merges predecessor blocks subject to contiguity, width, dependency-region boundaries, and acyclic graph contraction. Scoring each fused target by Eq.~\eqref{eq:proxy} and scheduling by Eq.~\eqref{eq:schedule} gives $\widehat T(x)$, incorporating the combined effects of layout and routing on targets and execution.

Population updates use these fitness values. Tournament selection chooses parents; numerical crossover acts on the priority and routing-weight vectors, while order crossover preserves valid layout permutations. Mutation includes numerical perturbations, layout swaps, and changes to the parent's critical path or fusion boundaries. Critical-path mutation adjusts the priorities, qubit positions, or routing weights associated with that path. Fusion-boundary mutation identifies neighboring blocks whose union is contiguous and within the width limit, then exchanges the priorities of source operations with no dependency between them to change the subsequent decoding order.

Each offspring is routed, fused, and scheduled again. Parents and offspring are ranked jointly by $\widehat T(x)$, retaining lower-scoring individuals. At termination, the selected fused targets and dependency graph pass to QOC.

\section{Methodology}
\label{sec:methodology}

The evaluation addresses four questions:
\begin{enumerate}
 \item[RQ1.] How much does SCOPE reduce QOC critical-path duration relative to the adapted EPOC reference?
 \item[RQ2.] Do surrogate and final QOC durations give consistent assessments of the relative gains between EPOC and SCOPE?
 \item[RQ3.] How do search results vary across random seeds for the same input?
 \item[RQ4.] How does final pulse latency change when gate count or gate-level depth replaces the time surrogate as search guidance?
\end{enumerate}

\subsection{Workloads and Experimental Settings}
\label{sec:experiment_settings}

The 13 fixed inputs cover quantum Fourier transforms (QFT: 6, 10, 15, 20 qubits), combinatorial optimization (QAOA: 6, 10, 15, 20), Hamiltonian simulation (Heisenberg: 6, 10, 15), and linear-system solving (HHL: 6, 10). Each instance has eight SCOPE seeds and one EPOC~\cite{Cheng2025EPOC} result, totaling 117 compilations. EPOC is adapted for this comparison.

SCOPE uses the same search and stopping criteria across instances. Section~\ref{sec:surrogate} specifies the control model; Appendix~\ref{app:reproduction} gives circuit generation, physical settings, and data aggregation.

Every block has a process-fidelity threshold of $0.9999$, with $F_b=|\operatorname{Tr}(U_b^\dagger\widetilde U_b)|^2/d_b^2$ for target $U_b$, propagated unitary $\widetilde U_b$ after frame restoration, and Hilbert-space dimension $d_b$. The numerical lower bound on $F_b$, accounting for integration error, must meet this threshold before the pulse duration enters the final schedule.

\subsection{Timing Metrics and Statistics}

For instance $c$, subscripts $\mathrm E$ and $\mathrm S$ denote EPOC and SCOPE. Let $T_{c,\mathrm E}$ be the reference duration and $T_{c,s}$ the duration of the $s$th SCOPE result. We report the arithmetic mean, sample standard deviation, and range over the eight seeds, and define the reduction as
\begin{equation}
 \overline T_{c,\mathrm S}=\frac18\sum_{s=1}^{8}T_{c,s},\qquad
 r_c=100\left(1-\frac{\overline T_{c,\mathrm S}}{T_{c,\mathrm E}}\right)\%.
 \label{eq:reduction}
\end{equation}
We average the reductions with equal weight across instances, $\overline r=\sum_{c=1}^{13}r_c/13$, and also report the geometric mean speedup,
\begin{equation}
 S_{\mathrm{geo}}=\exp\left[\frac1{13}\sum_{c=1}^{13}
 \log\frac{T_{c,\mathrm E}}{\overline T_{c,\mathrm S}}\right].
 \label{eq:speedup}
\end{equation}
Both aggregates weight instances equally. Durations are expressed in ns.

Consistency uses 13 paired EPOC--SCOPE comparisons. Equation~\eqref{eq:reduction} gives surrogate and QOC reductions, each normalized by EPOC for the same instance and metric, with SCOPE represented by its eight-run mean. Directional agreement counts pairs favoring the same method; the equally weighted mean absolute difference (percentage points) measures gain agreement; Pearson correlation measures their association across instances.

\subsection{Search-Guidance Ablation}
\label{sec:ablation_setup}

On QFT-6, we replace time-surrogate guidance with physical gate-count or gate-level depth guidance, using eight random seeds per arm. The time-surrogate arm reuses the eight QFT-6 results in Table~\ref{tab:qoc}. All three arms share the input, population size, seed set, routing, terminal layout restoration, and greedy fusion; each stops after $500$ consecutive generations without improvement in the best objective value. The replacement covers both fitness evaluation and objective-dependent mutation-region selection. Gate-count guidance selects regions over all logical operations, whereas depth guidance uses the gate-level critical path and its neighborhood. Final QOC durations use the local block threshold in Section~\ref{sec:experiment_settings} and are scheduled by Eq.~\eqref{eq:schedule}. Appendix~\ref{app:ablation} defines gate count and depth and describes the run-level records.

\section{Evaluation}

\label{sec:evaluation}

\subsection{Circuit Execution Latency and Search Variation (RQ1 and RQ3)}

SCOPE reduces mean final QOC critical-path duration by 42.50\% relative to adapted EPOC, weighting instances equally (Table~\ref{tab:qoc}). All 13 instances benefit: reductions range from 7.15\% to 82.71\%, with a median of 33.89\% and a geometric mean speedup of 2.00. Durations use numerically verified local pulses.

\begin{table}[htbp]
\centering
\small
\caption{Circuit critical-path durations computed from numerically verified local QOC pulse durations. SCOPE is run with eight random seeds per instance. SD denotes the sample standard deviation.}
\label{tab:qoc}
\setlength{\tabcolsep}{4pt}
\begin{tabular}{lrrrr}
\toprule
Circuit & EPOC~\cite{Cheng2025EPOC} / ns & SCOPE mean $\pm$ SD / ns & SCOPE range / ns & Reduction \\
\midrule
QFT-6 & 359.00 & 268.38 $\pm$ 22.18 & 249.00--295.50 & 25.24\% \\
QFT-10 & 991.00 & 655.25 $\pm$ 25.02 & 615.00--690.00 & 33.88\% \\
QFT-15 & 1704.00 & 1126.44 $\pm$ 24.91 & 1092.00--1173.00 & 33.89\% \\
QFT-20 & 2589.50 & 1585.69 $\pm$ 51.33 & 1526.50--1690.50 & 38.76\% \\
QAOA-6 & 718.00 & 639.06 $\pm$ 10.30 & 633.50--656.00 & 10.99\% \\
QAOA-10 & 1299.50 & 1206.63 $\pm$ 45.28 & 1127.50--1270.50 & 7.15\% \\
QAOA-15 & 2383.00 & 1828.75 $\pm$ 50.32 & 1756.50--1896.00 & 23.26\% \\
QAOA-20 & 3216.00 & 2824.13 $\pm$ 77.44 & 2710.50--2922.50 & 12.19\% \\
Heisenberg-6 & 2570.00 & 626.63 $\pm$ 1.83 & 623.00--629.00 & 75.62\% \\
Heisenberg-10 & 5448.50 & 1155.31 $\pm$ 30.10 & 1117.50--1178.00 & 78.80\% \\
Heisenberg-15 & 10597.00 & 1832.44 $\pm$ 9.18 & 1810.50--1838.00 & 82.71\% \\
HHL-6 & 1295.00 & 442.69 $\pm$ 10.46 & 437.00--462.00 & 65.82\% \\
HHL-10 & 9234.00 & 3310.94 $\pm$ 45.64 & 3258.50--3400.50 & 64.14\% \\
\bottomrule
\end{tabular}
\end{table}

\begin{figure}[!htbp]
\centering
\includegraphics[width=\linewidth]{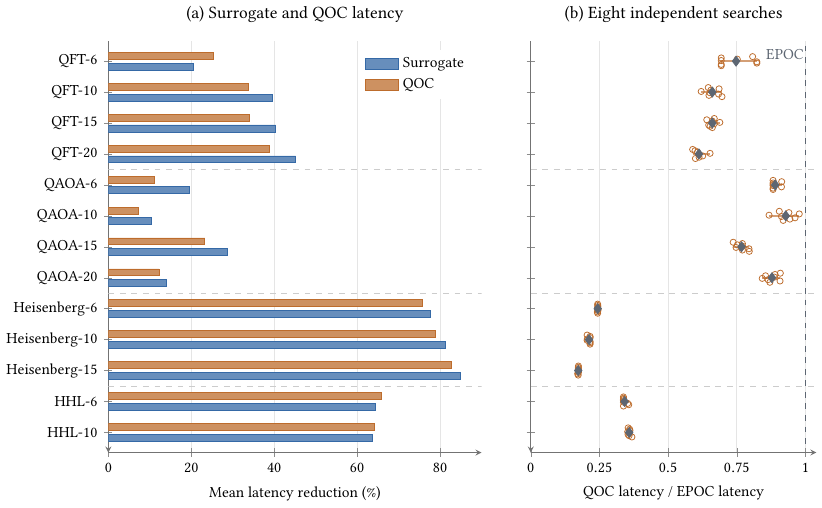}
\caption{Latency relative to same-instance EPOC. (a) Mean surrogate and QOC reductions. (b) All 104 SCOPE QOC results: circles mark runs, diamonds eight-run means, segments minimum--maximum ranges, and the dashed line EPOC. Vertical offsets separate coincident runs.}
\label{fig:reduction}
\Description{The left panel compares mean surrogate and QOC reductions for thirteen instances, with larger gains for Heisenberg and HHL. The right panel plots all 104 SCOPE QOC durations relative to the corresponding EPOC result. Every point lies to the left of the baseline ratio of one.}
\end{figure}

Figure~\ref{fig:reduction} compares mean reductions and individual-run durations relative to EPOC on the same instance.

The largest gains occur in Heisenberg and HHL, with family means of 79.04\% and 64.98\%, respectively. The three Heisenberg reductions are 75.62\%, 78.80\%, and 82.71\%; the two HHL reductions are 65.82\% and 64.14\%.

The benefit also extends to QFT and QAOA. QFT has an equally weighted mean reduction of 32.95\%, increasing from 25.24\% to 38.76\% across the observed sizes. QAOA averages 13.40\%, with reductions of 23.26\% for QAOA-15 and 12.19\% for QAOA-20. Thus, all four families benefit, while the gain's dependence on size varies by family.

All 104 SCOPE durations in Figure~\ref{fig:reduction}(b) are below EPOC, so the benefit extends from the instance means to every search result. On QAOA-10, the mean is $1206.63\pm45.28$ ns and the maximum is $1270.50$ ns, both below EPOC's $1299.50$ ns.

\begin{figure}[!htbp]
\centering
\includegraphics[width=\linewidth]{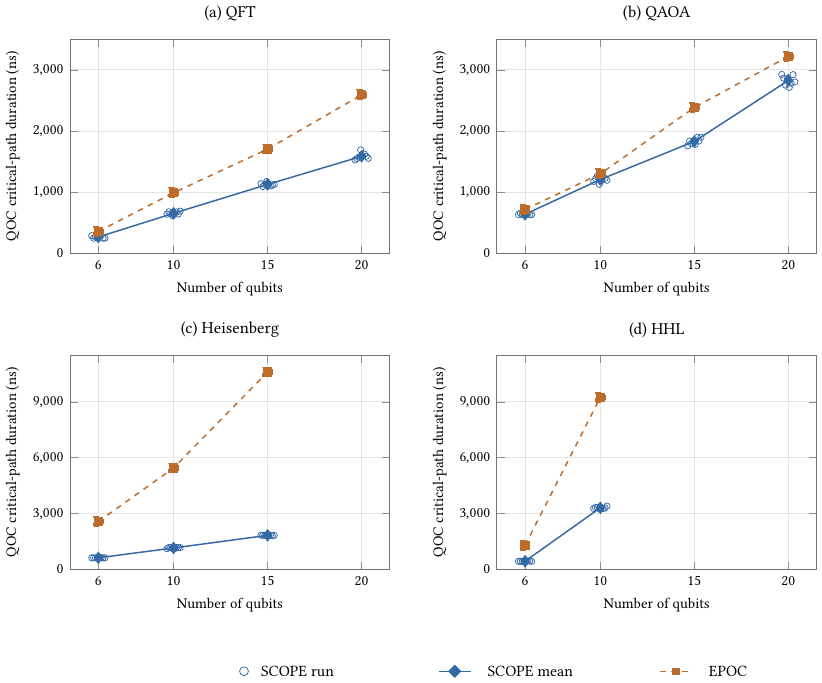}
\caption{QOC duration by circuit size. Squares mark EPOC; diamonds eight-run SCOPE means; circles individual runs; segments minimum--maximum ranges. Lines connect observed inputs, with horizontal offsets for coincident runs. Panels in each row share a vertical scale.}
\label{fig:scaling}
\Description{Four panels plot QOC critical-path duration against circuit size for QFT, QAOA, Heisenberg, and HHL. SCOPE means and individual runs are below EPOC at each observed instance. Heisenberg and HHL exhibit larger absolute timing gaps.}
\end{figure}

Figure~\ref{fig:scaling} shows the absolute timing scale, with lines connecting fixed inputs at each size. SCOPE and EPOC durations both grow with size for QFT and QAOA, while Heisenberg and HHL show larger absolute gaps.

\subsection{Agreement between Surrogate and QOC Gains (RQ2)}

\label{sec:comparison_consistency}

\begin{figure}[!htbp]
\centering
\includegraphics[width=\linewidth]{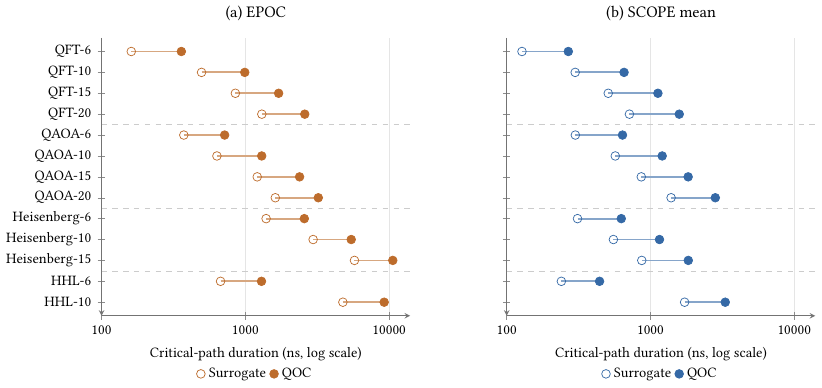}
\caption{Surrogate and QOC durations: segments connect surrogate scores (open circles) to QOC durations (solid circles) for EPOC (left) and eight-run SCOPE means (right). Both panels use the same logarithmic time axis.}
\label{fig:score_duration}
\Description{Two dot-and-segment plots show EPOC and SCOPE on a shared logarithmic time axis. A segment connects the surrogate score and QOC duration of each circuit instance. The QOC duration exceeds the surrogate score for every instance.}
\end{figure}

Table~\ref{tab:proxy} and Figure~\ref{fig:score_duration} compare EPOC and SCOPE surrogate scores with QOC durations for the same compilation results.

\begin{table}[htbp]
\centering
\small
\caption{Surrogate critical-path durations by instance. SCOPE results report the mean and sample standard deviation (SD) over eight searches; reductions are relative to EPOC for the same instance.}
\label{tab:proxy}
\setlength{\tabcolsep}{4pt}
\begin{tabular}{lrrr}
\toprule
Circuit & EPOC~\cite{Cheng2025EPOC} / ns & SCOPE surrogate mean $\pm$ SD / ns & Reduction \\
\midrule
QFT-6 & 161.00 & 127.81 $\pm$ 5.81 & 20.61\% \\
QFT-10 & 496.00 & 299.63 $\pm$ 9.61 & 39.59\% \\
QFT-15 & 852.50 & 508.63 $\pm$ 10.08 & 40.34\% \\
QFT-20 & 1302.00 & 714.00 $\pm$ 22.05 & 45.16\% \\
QAOA-6 & 373.50 & 300.38 $\pm$ 9.28 & 19.58\% \\
QAOA-10 & 634.50 & 569.31 $\pm$ 16.33 & 10.27\% \\
QAOA-15 & 1209.50 & 863.44 $\pm$ 23.12 & 28.61\% \\
QAOA-20 & 1616.00 & 1391.44 $\pm$ 41.86 & 13.90\% \\
Heisenberg-6 & 1394.50 & 311.00 $\pm$ 0.00 & 77.70\% \\
Heisenberg-10 & 2961.00 & 552.44 $\pm$ 9.06 & 81.34\% \\
Heisenberg-15 & 5737.50 & 870.06 $\pm$ 2.65 & 84.84\% \\
HHL-6 & 672.50 & 240.13 $\pm$ 0.23 & 64.29\% \\
HHL-10 & 4766.50 & 1727.56 $\pm$ 28.50 & 63.76\% \\
\bottomrule
\end{tabular}
\end{table}

For QFT-10, the SCOPE surrogate mean is $299.63$ ns and the QOC mean is $655.25$ ns; Heisenberg-15 gives $870.06$ ns and $1832.44$ ns, respectively. These absolute gaps reflect the different roles of the two evaluations: the score extracts necessary timing constraints, whereas QOC constructs complete pulses satisfying the control conditions. The comparison between methods therefore depends on whether these gaps change their relative gains.

\begin{figure}[!htbp]
\centering
\includegraphics[width=\linewidth]{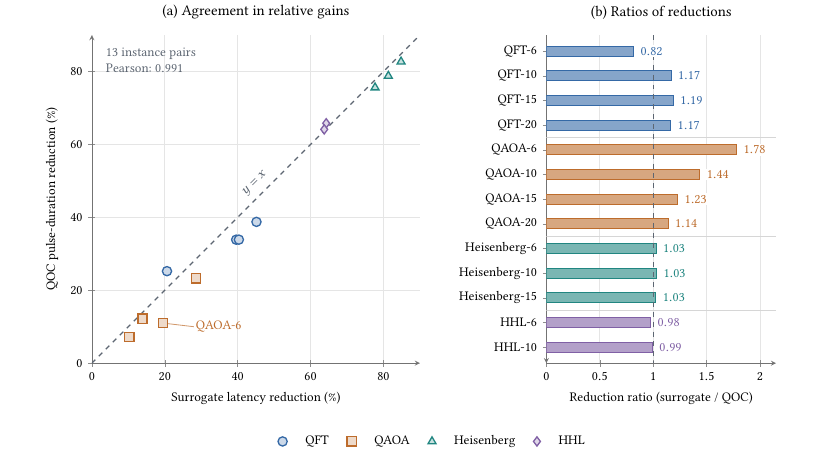}
\caption{EPOC--SCOPE gain agreement, comparing SCOPE eight-run means with EPOC. (a) Surrogate and QOC reductions, with equality line $y=x$. (b) Surrogate reduction divided by QOC reduction, with equality line $1$. Colors identify circuit families.}
\label{fig:consistency}
\Description{The scatter plot places surrogate reduction on the horizontal axis and QOC reduction on the vertical axis for thirteen instances. All points have positive coordinates and broadly follow the identity line. Horizontal bars show the ratio of these reductions; QAOA-6 is approximately 1.78, while Heisenberg and HHL are near one.}
\end{figure}

Figure~\ref{fig:consistency} tests this agreement by pairing surrogate and QOC reductions relative to same-instance EPOC and plotting their ratios.

Both evaluations favor SCOPE over EPOC in all 13 instances, giving directional agreement in $13/13$ comparisons. The two sets of reductions have a Pearson correlation of $0.991$ and a mean absolute difference of $3.89$ percentage points. In Figure~\ref{fig:consistency}(b), all three Heisenberg ratios round to $1.03$, while HHL-6 and HHL-10 give $0.98$ and $0.99$, indicating close agreement in their gains. QAOA-6 has the largest difference: surrogate and QOC reductions are 19.58\% and 10.99\%, a ratio of $1.78$ and a difference of $8.58$ percentage points. These comparisons show agreement on the direction of improvement, with final QOC durations determining the gain magnitude.

\subsection{Effect of Search Guidance on Pulse Duration (RQ4)}
\label{sec:ablation}

Time-surrogate guidance gives the shortest final QOC critical paths in the QFT-6 ablation (Table~\ref{tab:ablation}). Its mean QOC duration is $268.38$ ns, 22.21\% below the gate-count mean of $345.00$ ns and 21.01\% below the depth mean of $339.75$ ns. All eight durations ($249.00$--$295.50$ ns) are below the other arms' minima of $317.50$ ns and $305.00$ ns.

\begin{table}[H]
\centering
\small
\caption{QFT-6 search-guidance ablation. Each strategy has eight runs; the time-surrogate reference reuses the main-experiment results. Gate count and depth are means; surrogate and QOC durations are means $\pm$ sample SD.}
\label{tab:ablation}
\setlength{\tabcolsep}{4pt}
\begin{tabular}{lrrrr}
\toprule
Search guidance & Gate count & Gate depth & Surrogate / ns & QOC / ns \\
\midrule
Time surrogate & 162.75 & 84.63 & 127.81 $\pm$ 5.81 & 268.38 $\pm$ 22.18 \\
Gate count & 150.00 & 101.38 & 184.19 $\pm$ 14.17 & 345.00 $\pm$ 19.01 \\
Gate-level depth & 162.00 & 69.00 & 188.13 $\pm$ 10.43 & 339.75 $\pm$ 19.62 \\
\bottomrule
\end{tabular}
\end{table}

The structural results explain the choices made by the different objectives. Gate-count guidance reduces the mean count from $162.75$ to $150.00$, or 7.83\%; depth guidance reduces mean depth from $84.625$ to $69.00$, or 18.46\%. Time-surrogate guidance selects circuits according to their fused execution schedules, retaining more gates or layers while achieving shorter QOC duration. Section~\ref{sec:discussion} relates this outcome to the physical timing information and fused targets used by the score.

\section{Discussion}
\label{sec:discussion}

\subsection{Effectiveness of the Time Surrogate}

The QFT-6 ablation holds the input, compilation pipeline, and stopping rule fixed while replacing fitness and its corresponding mutation-region selection (Section~\ref{sec:ablation_setup}). The resulting QOC durations test whether physical timing information improves the implementations selected by this search.

Table~\ref{tab:ablation} gives direct evidence of that effect. Time-surrogate guidance produces a mean QOC duration of $268.38$ ns, compared with $345.00$ ns for gate-count guidance and $339.75$ ns for depth guidance, reductions of $22.21\%$ and $21.01\%$. All eight durations ($249.00$--$295.50$ ns) are below the other arms' minima of $317.50$ and $305.00$ ns. Evaluating all outputs with the same surrogate also gives the time-guided arm the lowest mean score, $127.81$ ns versus $184.19$ and $188.13$ ns.

The useful information enters through the fused target. Gate count describes an operation sequence, whereas joint control acts on its product unitary. For any positive integer $m$, the identity $R_x(\theta)=[R_x(\theta/m)]^m$ shows that different counts can represent the same target; the XY example in Figure~\ref{fig:xy_bound} shows that equal gate types and counts can require different control times. The Schmidt angle measures the target change requiring cross-cut coupling, and the coupling bound limits its rate. Their ratio therefore supplies a physical time constraint that changes with the target. The trace-angle and waveform terms extend this constraint to local rotations and admissible pulses.

Search must also account for where these timing requirements occur. Each candidate's block scores become node weights in the fused dependency graph, and the longest weighted path determines fitness. Layout, routing, and gate order can thus improve fitness by changing targets, dependencies, or both. In the ablation, every gate-count-guided result contains $150$ gates, yet their QOC durations span $317.50$--$368.50$ ns; every depth-guided result has depth $69$, yet durations span $305.00$--$361.50$ ns. These variations show that equal structural scores do not determine QOC duration. The comparison among guidance strategies then tests the benefit of using a target-dependent timing score.

An effective latency objective must both resolve this timing variation and use it to select better implementations. The repeated structural scores establish the first requirement: fitness evaluation based on those scores assigns equal values to these outcomes with different pulse times. The separation of the QOC ranges then tests the second requirement in the shared pipeline. Every time-guided result is shorter than every result obtained with the two structural objectives. The ablation thus connects information retained by the physical score to the implementations selected when that information guides fitness and mutation-region selection.

The objectives explain the different structural choices. Relative to the time-surrogate arm, gate-count guidance reduces the mean count by $7.83\%$, and depth guidance reduces mean depth by $18.46\%$. The time surrogate instead evaluates the fused execution schedule, allowing search to retain more gates or layers when the surrogate critical path is shorter. The measured QOC gains validate this joint use of physical timing information for fitness evaluation and mutation-region selection.

The EPOC--SCOPE results provide complementary evidence: surrogate and QOC reductions agree in direction across all 13 instances, with a mean absolute difference of $3.89$ percentage points (Section~\ref{sec:comparison_consistency}). The ablation tests using the proxy to guide search; the framework comparison assesses whether its relative gains are retained after pulse optimization.

\subsection{The Timing Information Retained by the Surrogate}

The lower bound responds to the evolution required by a target. Local operations preserve its cut Schmidt spectrum, so the cut angle measures change that must be produced by the bounded coupling. In the amplitude-only XY example of Figure~\ref{fig:xy_bound}, the cut bound and the feasible duration $T_\varphi=\varphi/\bar g$ have the same leading dependence on interaction angle: their ratio approaches one as $\varphi\to0$ (Appendix~\ref{app:xy}). The bound therefore captures a target-dependent control-time scale even when gate type and count stay fixed.

The circuit score carries this information into scheduling. Waveform padding and grid rounding are monotone in the raw block bound. For a fixed dependency graph, the longest weighted path is likewise monotone in block durations. A candidate can therefore improve its score by reducing block time bounds or changing dependencies to allow more parallel execution. The shared physical constraints give these choices a common time scale, while QOC determines their realized durations.

To examine how the gains survive this realization, define $\gamma_m=T_m^{\mathrm{QOC}}/\widehat T_m$ for method $m\in\{\mathrm E,\mathrm S\}$ on one instance, using the eight-run mean for SCOPE. Then
\begin{equation}
 \frac{T_{\mathrm S}^{\mathrm{QOC}}}{T_{\mathrm E}^{\mathrm{QOC}}}
 =\kappa\frac{\widehat T_{\mathrm S}}{\widehat T_{\mathrm E}},
 \qquad \kappa=\frac{\gamma_{\mathrm S}}{\gamma_{\mathrm E}}.
 \label{eq:relative_gap}
\end{equation}
A common multiplicative gap cancels in the comparison: when $\kappa=1$, surrogate and QOC reductions coincide. More generally, the QOC gain remains positive when $\kappa<\widehat T_{\mathrm E}/\widehat T_{\mathrm S}$, so a larger surrogate advantage tolerates a larger difference between realization factors.

Writing the fractional reductions as $r_{\mathrm{sur}}=1-\widehat T_{\mathrm S}/\widehat T_{\mathrm E}$ and $r_{\mathrm{QOC}}=1-T_{\mathrm S}^{\mathrm{QOC}}/T_{\mathrm E}^{\mathrm{QOC}}$ makes this transmission explicit: $r_{\mathrm{QOC}}-r_{\mathrm{sur}}=(1-\kappa)(1-r_{\mathrm{sur}})$. The relative realization factor determines the direction of the change, while the remaining surrogate-time fraction determines its magnitude. Hence the same change in $\kappa$ has a smaller effect when the surrogate reduction is larger. This relation gives a direct interpretation of Figure~\ref{fig:consistency}: deviations from equality reflect how the two methods' realization factors differ on each instance.

Tables~\ref{tab:qoc} and~\ref{tab:proxy} illustrate both effects; across all 13 cases, $\kappa$ ranges from $0.942$ to $1.140$. On HHL-10, the QOC/surrogate factors are $1.937$ for EPOC and $1.917$ for SCOPE. Their ratio is $0.989$, and the gain changes from $63.76\%$ to $64.14\%$. On QAOA-6, the surrogate gain is $19.58\%$ and the realization-factor ratio is $\kappa\approx1.107$, below the threshold of $1.243$ for a positive QOC gain. The resulting QOC reduction is $10.99\%$. This decomposition quantifies how the relative gains change after QOC; the geometric bounds and scheduling rule supply the timing information used before QOC.

\subsection{Preservation of Search Gains after Pulse Optimization}
\label{sec:bound_discrimination}

For a general fused target, the cut and trace angles constrain the endpoint of an evolution, whereas a pulse must follow an admissible path to that endpoint. Taking the maximum of these constraints gives a necessary duration; attaining it also requires compatible evolution directions and waveform controls. The circuit score therefore retains mandatory timing requirements while leaving the control path to QOC. The ablation tests whether the resulting score favors implementations whose advantage survives that step.

An additive form of the same comparison uses the signed difference $e_x=T_x^{\mathrm{QOC}}-\widehat T_x$ between the exact-target score and the duration at the prescribed block threshold. It includes changes in block durations and in the critical path. For a lower-scoring result $x$ and a comparator $y$,
\begin{equation}
 T_y^{\mathrm{QOC}}-T_x^{\mathrm{QOC}}
 =(\widehat T_y-\widehat T_x)+(e_y-e_x).
 \label{eq:ranking_margin}
\end{equation}
The score advantage is retained when it exceeds $e_x-e_y$, and grows when $e_x<e_y$. The same relation applies to arm means, giving a direct way to interpret the ablation.

Relative to gate-count guidance, the time-surrogate arm has a $56.38$ ns mean score advantage and a $20.25$ ns smaller mean $e_x$. Evaluating Eq.~\eqref{eq:ranking_margin} with the unrounded means gives a $76.63$ ns QOC advantage. Against depth guidance, the corresponding terms are $60.31$ and $11.06$ ns, giving $71.38$ ns. The selected circuits thus combine shorter surrogate critical paths with smaller additional QOC durations. The ablation's observed reductions of $22.21\%$ and $21.01\%$ validate the joint time-surrogate guidance in this search pipeline.

Fusion statistics further connect the result to the execution units. Time-surrogate guidance produces $11.88$ blocks on average, compared with $13.25$ under gate-count guidance and $15.13$ under depth guidance. Every depth-guided output has $162$ gates and depth $69$, no greater than any time-surrogate output, yet all its QOC durations are longer. Gate layers therefore omit differences in the fused targets and their timing. The proposed surrogate evaluates these targets together with their block dependencies, matching the level at which pulses are constructed and scheduled.

\subsection{Limitations}
\label{sec:limitations}

This study considers static circuits of unitary operations. Dynamic circuits with mid-circuit measurement, reset, and classical feedforward offer a further setting to explore. Measurement outcomes select subsequent branches, making quantum evolution and classical decisions parts of the same execution process. A corresponding timing objective could account for measurement, reset, and feedforward waits together with pulse durations within each branch. For circuits with finitely many branches, the probability-weighted mean duration measures average cost across repeated runs, while the longest branch characterizes worst-case response time. These objectives address different execution requirements, extending the study of physical execution latency from fixed quantum schedules to conditional execution. Further investigation could examine the timing relationships between quantum operations and classical feedback, and the conditions under which pulse-duration gains translate into shorter program execution.

Fault-tolerant logical circuits offer another level of extension. Logical execution involves error-correction cycles, syndrome measurements, and decoding feedback; surface-code research also reports cycle time and decoder latency as distinct operational quantities~\cite{Google2025BelowThreshold}. Overlap between quantum operations and classical processing, together with dependencies that require feedback, determines the logical critical path. Future work could therefore study how physical operation durations relate to correction cycles, decoding feedback, and logical-operation completion times under a specified code and target logical reliability. The central question is when physical timing gains propagate to the logical level, and how compilation choices then affect the execution time of a complete fault-tolerant program. This setting extends the physical-latency objective to logical operations and provides a basis for studying the timing relationship between physical control and fault-tolerant execution.

\section{Conclusions and Future Work}
\label{sec:conclusion}

SCOPE connects structural search to physical pulse latency through operator Schmidt time bounds, local-rotation constraints, and waveform requirements. The bound is available from the target and admissible control resources before pulse construction, allowing candidate structures to be compared while they can still be changed. Critical-path scheduling translates these block requirements into the time of the complete dependency graph. The compiler can therefore act on both the control demands of the fused targets and their serial accumulation in the circuit, using a common objective for layout, operation-order, and routing choices, accounting for their coupled effects on execution time. QOC then realizes the selected candidates and determines their final durations.

Under the numerical control model, all 104 SCOPE results across 13 instances from four application families have shorter QOC critical-path durations than adapted EPOC, with an equally weighted mean reduction of 42.50\% and a geometric mean speedup of 2.00. Surrogate scores and QOC durations favor SCOPE in all 13 paired comparisons. Their agreement connects the timing criterion used to compare targets with the final execution outcome. The ablation then tests this guidance within the same compilation pipeline. In the shared QFT-6 pipeline, time guidance reduces mean QOC duration by 22.21\% relative to gate-count guidance and 21.01\% relative to depth guidance. Every time-guided result is shorter than every result in the other two arms, validating the physical time score's joint use in fitness evaluation and mutation-region selection.

As physical control capabilities evolve, quantum compilation needs timing objectives that connect target operations and execution dependencies to available resources. SCOPE establishes this connection before pulse generation, allowing structural choices to be compared through the physical resources available to implement them, and validates its benefit in final QOC execution latency.

\begin{acks}
This work is supported by the National Key Research and Development Program of China (Grant No.~2024YFB4504103). Numerical simulations were performed on the SongShan supercomputer at the National Supercomputing Center in Zhengzhou. This work is also supported by Jiangsu Province Engineering Research Center of IntelliSense Technology and System.
\end{acks}

\section*{Code and Data Availability}
\makeatletter
\if@ACM@anonymous\else
SCOPE source code is hosted at \url{https://github.com/huang4210/SCOPE}.\par
\fi
\makeatother
The accompanying supplementary material provides the input circuits, per-run timing and ablation results, and a Python script for reproducing the reported statistics. Appendix~\ref{app:reproduction} specifies the circuit-generation rules and the physical and search settings.

\clearpage
\appendix
\section{Derivation of Control-Time Bounds}
\label{app:bounds}

We derive the Schmidt, trace-angle, and waveform-envelope time constraints for the control model in Eq.~\eqref{eq:hamiltonian}.

\subsection{Operator Geometry and Product Overlap}
\label{app:geometry}
Represent a unitary by the normalized vector $|R\rangle\!\rangle=\operatorname{vec}(R)/\sqrt d$ and use the angular distance $D(\psi,\phi)=\arccos|\langle\psi,\phi\rangle|$. For a partition $A|\bar A$, let $\mathcal S_A$ contain all normalized product vectors in the two operator subspaces. The relation between the Schmidt decomposition and maximal product-state overlap gives~\cite{Wei2003Geometric}
\begin{equation}
 \max_{\phi\in\mathcal S_A}|\langle\phi|R\rangle\!\rangle|
 =\frac{s_1^{(A)}}{\sqrt d}=\sqrt{\lambda_A},\qquad
 \theta_A=\min_{\phi\in\mathcal S_A}D(|R\rangle\!\rangle,\phi).
 \label{eq:geometric_demand}
\end{equation}
The product-operator set contains all product unitaries across the cut. Operations confined to either side preserve the Schmidt spectrum and hence $\theta_A$. Starting from the identity, producing a nonzero angle therefore requires cross-cut coupling.

\subsection{Control Resources and Evolution Speed}
\label{app:resources}
Write $K(t)=H(t)/\hbar$. Let $W(t)$ evolve from the identity and define its normalized operator vector $|w(t)\rangle\!\rangle=\operatorname{vec}(W(t))/\sqrt d$. Its projective-space speed under the angular distance above is~\cite{Impens2025GeometricControl,Anandan1990Geometry}
\begin{equation}
 v_{\mathrm{FS}}(t)^2
 =\langle\!\langle\dot w|\dot w\rangle\!\rangle
  -|\langle\!\langle w|\dot w\rangle\!\rangle|^2
 =\frac{\operatorname{Tr}[K(t)^2]}{d}
  -\left(\frac{\operatorname{Tr}K(t)}d\right)^2.
 \label{eq:operator_speed}
\end{equation}
Unitarity makes this speed depend only on the instantaneous generator.

Separate the generator into within-side terms $K_A+K_{\bar A}$ and a cross-cut term $K_{\partial A}$. The former generate $V_A\otimes V_{\bar A}$. In the interaction picture $W_I=(V_A\otimes V_{\bar A})^\dagger W$, only a unitary conjugate of $K_{\partial A}$ drives evolution. This transformation preserves the target's $\theta_A$ and the coupling's Hilbert--Schmidt norm. Orthogonality of Pauli strings therefore gives, for any exact implementation of $R$,
\begin{equation}
 \theta_A(R)\le D(|I\rangle\!\rangle,|W_I(T)\rangle\!\rangle)
 \le\int_0^T\sqrt{\frac12\sum_{e\in\delta A}g_e(t)^2}\,dt
 \le T v_A.
 \label{eq:cut_speed_bound}
\end{equation}
The inequalities respectively use $I\in\mathcal S_A$, endpoint distance bounded by path length, and the coupling-amplitude limits. Thus, $T\ge\theta_A/v_A$ for every cut; taking the maximum enforces all cut constraints.

Product unitaries have zero cut angles but can have nonzero trace angle $\theta_{\mathrm{all}}$. Equation~\eqref{eq:operator_speed} and the componentwise amplitude bounds also constrain their local-rotation time:
\begin{equation}
 v_{\mathrm{FS}}(t)^2
 =\frac14\sum_i\bigl(u_{ix}(t)^2+u_{iy}(t)^2\bigr)
  +\frac12\sum_e g_e(t)^2
 \le v_{\mathrm{all}}^2,
 \qquad T\ge\frac{\theta_{\mathrm{all}}}{v_{\mathrm{all}}}.
 \label{eq:global_speed_bound}
\end{equation}
Together, the constraints give $T\ge\widehat t_{\mathrm{raw}}$. Minimizing the trace angle over the allowed frames in Eq.~\eqref{eq:frame} preserves the cut angles and accounts for zero-duration frame updates.

\subsection{Analytic Relation for the XY Family}
\label{app:xy}

For the XY gate family, the cut bound can be compared analytically with a feasible control duration. Let $R_\varphi=\exp[-i\varphi(XX+YY)/2]$, with $0<\varphi\le\pi/2$. Its expansion is $c^2II+s^2ZZ-ics(XX+YY)$, where $c=\cos(\varphi/2)$ and $s=\sin(\varphi/2)$, giving the largest normalized Schmidt coefficient $c^2$. Under amplitude constraints alone, a constant coupling $\bar g$ produces the feasible duration $T_\varphi=\varphi/\bar g$, while
\begin{equation}
 \frac{\theta_A/v_A}{T_\varphi}
 =\frac{\sqrt2\,\arccos[\cos^2(\varphi/2)]}{\varphi}
 \longrightarrow1\quad(\varphi\to0).
 \label{eq:xy_example}
\end{equation}
The ratio is $2\sqrt2/3\approx0.943$ at $\varphi=\pi/2$ and approaches one for small angles: the bound is asymptotically tight relative to this feasible duration (Fig.~\ref{fig:xy_bound}).

\subsection{Waveform Envelope and Time Grid}
\label{app:waveform}
The channels share a nonnegative, normalized FIR kernel, zero endpoint padding, and piecewise-linear interpolation. Setting normalized raw nodes to one defines $e_T(t)$, which bounds the absolute value of every normalized component by positivity. For $N$ raw nodes spaced by $\Delta t$, convolution preserves their sum, and zero endpoints make the integral equal to that sum times the spacing:
\begin{equation}
 0\le e_T(t)\le1,\qquad
 \int_0^T e_T(t)\,dt=N\Delta t=T-\tau_{\mathrm{pad}}.
 \label{eq:envelope_area}
\end{equation}
Normalization makes the envelope area a time, with $\tau_{\mathrm{pad}}$ its deficit from a rectangular envelope. Replacing the speed bounds in Eqs.~\eqref{eq:cut_speed_bound} and~\eqref{eq:global_speed_bound} by $e_T(t)v_A$ and $e_T(t)v_{\mathrm{all}}$ gives $\widehat t_{\mathrm{raw}}\le T-\tau_{\mathrm{pad}}$. Incorporating $\tau_{\min}$ and the time grid yields Eq.~\eqref{eq:proxy}.

\subsection{Circuit-Level Bounds}
\label{app:circuit_bounds}
For a fixed block partition and physical dependency graph, the critical-path duration is monotone in every block duration. Substituting the exact-target block bounds into Eq.~\eqref{eq:schedule} therefore yields a circuit-duration bound for that partitioned control problem.

\section{Reproduction Settings and Data}
\label{app:reproduction}

\subsection{Circuit Generation}
\label{app:circuit_generation}

Both methods use identical fixed inputs containing only $R_X$, $R_Z$, and CX operations, with $q_0$ the most significant bit. Table~\ref{tab:workloads} gives their sizes and gate counts before routing.

\begin{table}[htbp]
\centering
\caption{Fixed compilation inputs before physical routing. Gate counts refer to their common $R_X/R_Z/\mathrm{CX}$ representation; values in each row follow the listed qubit counts.}
\label{tab:workloads}
\begin{tabular}{lll}
\toprule
Circuit family & Qubits & Native gate counts \\
\midrule
QFT & 6, 10, 15, 20 & 102, 270, 591, 1040 \\
QAOA & 6, 10, 15, 20 & 117, 195, 288, 390 \\
Heisenberg & 6, 10, 15 & 1188, 1988, 2988 \\
HHL & 6, 10 & 246, 1071 \\
\bottomrule
\end{tabular}
\end{table}

\paragraph{QFT}
The QFT inputs implement the full transform $F_{y,x}=2^{-n/2}\exp(2\pi ixy/2^n)$. For each target $j$, a Hadamard gate is followed by controlled phases from qubits $k>j$ with angles $\pi/2^{k-j}$. All controlled phases are retained. Final swaps between $j$ and $n-1-j$, for $0\le j<\lfloor n/2\rfloor$, implement bit reversal.

\paragraph{QAOA}
For the same four sizes, the unweighted MaxCut graph contains cycle edges $(j,(j+1)\bmod n)$ and matching edges $(j,j+\lfloor n/2\rfloor)$ for $0\le j<\lfloor n/2\rfloor$. Starting from $H^{\otimes n}$, three layers apply $\exp(-i\gamma_\ell H_C)$ followed by $\exp(-i\beta_\ell H_B)$, where $H_C=\sum_{(u,v)\in E}(I-Z_uZ_v)/2$ and $H_B=\sum_jX_j$. The fixed parameters are $\boldsymbol\gamma/\pi=(1/7,1/5,1/3)$ and $\boldsymbol\beta/\pi=(1/10,1/8,1/6)$. Within each cost layer, the cycle edges are processed in vertex order, followed by the matching edges in increasing $j$. This defines a fixed compilation target at each size.

\paragraph{Heisenberg}
For $n\in\{6,10,15\}$, the periodic model has $H_{\mathrm H}=\sum_{j=0}^{n-1}(X_jX_{j+1}+Y_jY_{j+1}+Z_jZ_{j+1})$, with indices modulo $n$, unit coupling, and zero external field. Evolution over model time $t=1$ is approximated by four symmetric second-order Trotter steps of size $1/4$. Terms are ordered by edges $(0,1),\ldots,(n-2,n-1),(n-1,0)$ and by $XX$, $YY$, $ZZ$ within each edge. A step applies all but the last term for half a step, the last term for a full step, and the preceding terms in reverse order for half a step. Consecutive steps have no inserted barriers. The model time specifies the simulated evolution; physical pulse duration is measured separately in ns.

\paragraph{HHL}
The two inputs encode the Dirichlet Poisson problem $-u''(x)=1$ on $(0,1)$, with $u(0)=u(1)=0$ and $N=4$ or $8$ interior points. Define $L_N=\operatorname{tridiag}(-1,2,-1)$ and $A=(N+1)^2L_N$, with a normalized uniform right-hand side. The system, phase, and flag registers contain $(2,3,1)$ or $(3,6,1)$ qubits, giving total sizes of 6 and 10. Construction uses the finite discrete-sine eigenbasis, phase estimation for $\exp(2\pi iL_N/8)$, controlled reciprocal rotations with $C=\lambda_{\min}(L_N)$ and zero amplitude at phase label zero, inverse phase estimation, and restoration of the sine basis. The canonical ordering places the flag at $q_0$, followed by the phase and system registers. Within each register, increasing wire index corresponds to decreasing bit significance.

\subsection{Physical Settings}

Surrogate scoring and QOC use the same physical settings on a homogeneous nearest-neighbor chain. The local I/Q components are bounded independently by $|u_{ix}|,|u_{iy}|\le\Omega$, with $\Omega/(2\pi)=100\,\mathrm{MHz}$, and signed XY couplings satisfy $|g_e|\le\bar g$, with $\bar g/(2\pi)=20\,\mathrm{MHz}$. The corresponding angular-frequency bounds in Eq.~\eqref{eq:hamiltonian} are $\Omega\approx0.628319\,\mathrm{rad/ns}$ and $\bar g\approx0.125664\,\mathrm{rad/ns}$; the XY generator is $(XX+YY)/2$. Drift and inter-block couplings during parallel execution are zero, and each fused block spans at most three consecutive physical qubits. The time grid is $\Delta t=0.5\,\mathrm{ns}$. Raw control nodes undergo full FIR convolution with kernel $(1,4,6,4,1)/16$, followed by two zero nodes at each end and piecewise-linear interpolation. The envelope-area relation in Appendix~\ref{app:waveform} gives $\tau_{\mathrm{pad}}=3.5\,\mathrm{ns}$ and a minimum nonzero waveform duration of $\tau_{\min}=4\,\mathrm{ns}$.

Each local block uses the process-fidelity threshold $0.9999$ defined in Section~\ref{sec:experiment_settings}. Final circuit latency is obtained by rescheduling with the numerically verified block durations.

\subsection{Experimental Records and Statistical Aggregation}

The 117 main-experiment records contain one EPOC result and eight SCOPE results per instance. Within-instance means, sample standard deviations, ranges, reductions, and consistency measures follow Section~\ref{sec:methodology}, using the corresponding EPOC result for each timing metric.

\subsection{QFT-6 Ablation Setup and Data}
\label{app:ablation}

The ablation uses the QFT-6 input in Table~\ref{tab:workloads} and the physical settings above. Physical gate count includes $R_X$, $R_Z$, and CX operations after routing and terminal layout restoration, with each SWAP decomposed into three CX gates. Gate-level depth is the longest unit-weight path on the physical-qubit dependency graph.

The surrogate arm reuses eight main-experiment endpoints and QOC results; the other arms repeat search and pulse optimization. All share population size, seeds, and deterministic tie rules. Objective-dependent selection, mutation, and stopping follow Section~\ref{sec:ablation_setup}, with termination after $500$ generations without improvement in the respective best objective.

The $24$ records contain gate count, depth, block count, and both durations; Table~\ref{tab:ablation} uses the main comparison's aggregation.

\end{document}